\documentclass[aps,twocolumn]{revtex4}%
\usepackage{amssymb}
\usepackage{amsmath}
\usepackage{graphicx}
\usepackage{amsfonts}%
\providecommand{\U}[1]{\protect\rule{.1in}{.1in}}

\begin{document}
\preprint{Final revision}
\title{Optimizing Anharmonicity in Nanoscale Weak Link Josephson Junction Oscillators}
\author{R. Vijay$^{1}$, J. D. Sau$^{2}$, Marvin L. Cohen$^{2}$, and I. Siddiqi$^{1}$}
\affiliation{$^{1}$Quantum Nanoelectronics Laboratory, Department of Physics, University of
California, Berkeley, CA 94720}
\affiliation{$^{2}$Department of Physics, University of California, Berkeley, CA 94720 and
Materials Sciences Division, Lawrence Berkeley National Laboratory, Berkeley,
CA 94720}

\begin{abstract}
Josephson tunnel junctions are widely used as nonlinear elements in
superconducting circuits such as low noise amplifiers and quantum bits.
However, microscopic defects in the oxide tunnel barrier can produce low and
high frequency noise which can potentially limit the coherence times and
quality factors of resonant circuits. Weak link Josephson junctions are an
attractive alternative provided that sufficient nonlinearity can be
engineered. We compute the current phase relation for superconducting weak
links, with dimensions comparable to the zero temperature coherence length,
connected to two and three dimensional superconducting electrodes. Our results
indicate that 50-100 nm long aluminum nanobridges connected with three
dimensional banks can be used to construct nonlinear oscillators for
bifurcation amplification. We also show that under static current bias, these
oscillators have a sufficiently anharmonic energy level structure to form a
qubit. Such weak link junctions thus present a practical new route for
realizing sensitive quantum circuits.

\end{abstract}
\volumenumber{number}
\issuenumber{number}
\eid{identifier}
\received[Received text]{date}

\revised[Revised text]{date}

\accepted[Accepted text]{date}

\published[Published text]{date}

\startpage{1}
\endpage{2}
\maketitle

Nonlinear oscillators with low loss are a basic building block of
superconducting quantum information processing circuits. They are used both as
\textquotedblleft artificial atoms\textquotedblright\ to realize quantum bits
\cite{Devoret2007}, and as the gain element in low noise amplifiers
\cite{Clerk2008} for quantum state measurement. The Josephson tunnel junction,
formed by separating two superconducting electrodes with a barrier such as a
thin oxide layer \cite{Josephson1964}, is the nonlinear circuit element
commonly used to construct these anharmonic oscillators. The current $I(t)$
and voltage $V(t)$ of the junction can be expressed in terms of $\delta(t)$,
the gauge-invariant phase difference, as $I(t)=I_{0}\sin\delta\left(
t\right)  $ and $V(t)=\varphi_{0}\ d\delta/dt$, where the parameter $I_{0}$ is
the junction critical current and $\varphi_{0}=\hbar/2e$ is the reduced flux
quantum. These equations parameterize a nonlinear inductance $L_{J}%
=\varphi_{0}/\left\{  I_{0}\cos(\delta)\right\}  $. This Josephson inductance
is shunted in parallel by a capacitance $C_{S}$ resulting from the geometric
capacitance of the junction and possibly an additional external capacitor,
thereby forming an electrical oscillator with an amplitude dependent plasma
oscillation frequency $\omega_{P}=1/\sqrt{L_{J}(I)C_{S}}$. In principle,
Josephson oscillators should exhibit a very large quality factor (Q) since at
temperatures $T$ well below the critical temperature $T_{c}$, dissipative
thermal quasiparticle production is exponentially suppressed. In practice,
microscopic defects in the tunnel barrier \cite{Simmonds2004} and in the shunt
capacitor dielectric \cite{Martinis2005} can reduce Q.

An alternative structure to the oxide barrier tunnel junction is a weak link
Josephson junction (see \cite{Likharev1979} for a review) where a small
metallic constriction with dimensions of order the superconducting coherence
length $\xi(T)$ bridges two large superconducting electrodes. This type of
junction also behaves as a nondissipative inductance but the strength of the
nonlinearity is a sensitive function of the weak link and electrode geometry.
In contrast to a tunnel junction which invariably has a sinusoidal current
phase relation (CPR) with $I\propto\sin(\delta)$, a weak link junction at
$T<<T_{c}$ can exhibit a wide range of CPRs ranging from $I\propto\delta
$---characteristic of a linear inductor---to a distorted sinusoid
$I\propto\cos(\delta/2)\tanh^{-1}[\sin(\delta/2)]$ for an idealized, short
constriction \cite{Kulik1975}. In this letter, we present calculated CPRs,
obtained via a numerical solution of the Usadel equations \cite{Usadel1970},
for thin, diffusive superconducting bridges contacted with two and three
dimensional banks for varying \ bridge length and width. We also compute the
dynamics of a weak link Josephson oscillator under ac excitation, and its
bound state energies. Our results indicate that with a 50-100nm long, 45nm
wide aluminum bridge with three dimensional banks, the oscillator is
sufficiently nonlinear to observe a stable dynamical bifurcation with a fixed
frequency microwave drive, and to obtain three quantized levels with $\sim3\%$
anharmonicity under dc bias. These results are obtained without putting
impractical constraints on the shunting capacitance, oscillator quality
factor, or current bias stability, and thus suggest a new route for realizing
superconducting qubits and amplifiers without tunnel junctions. Furthermore,
this junction geometry is suitable for coupling Josephson devices to other
solid state quantum systems such as nanomagnets and quantum dots.

We consider two canonical weak link geometries: a \textquotedblleft
Dayem\textquotedblright\ bridge where the banks are the same thickness as the
weak link and a \textquotedblleft variable thickness bridge\textquotedblright%
\ where the banks are significantly thicker than the weak link. Considerable
literature exists on both of these structures. As $T\longrightarrow T_{c}$,
the temperature dependent coherence length $\xi(T)$ diverges and mean field
methods, such as Ginzburg-Landau theory \cite{Ginzburg1950}, can be used to
compute the magnitude and phase of the superconducting order parameter along
the bridge and in the banks \cite{Hasselbach2002}. For qubits and low noise
amplifiers, we are interested in operating the Josephson junction at $T\ll
T_{c}$ to minimize quasiparticle loss. In this regime, linear response methods
based on Gorkov functions have been applied to short weak links with length
$L$ and width $W\ll\xi(T)$, and with rigid boundary conditions imposed at the
bridge ends --- Kulik-Omelyanchuk (K-O) theory \cite{Kulik1975}. However,
$10-30$ nm thick Al bridges with a mean free path $l\sim1$ nm
\cite{Siddiqi2002} have $\xi(0)\approx30$ nm, making the short limit difficult
to achieve with conventional e-beam lithography. We thus focus on weak links
with lateral dimensions comparable to $\xi(T=0)$ and large compared to $l$. In
the absence of closed form expressions for the CPR in this parameter regime,
we apply the Nambu-Gorkov formalism \cite{Gorkov1958, Nambu1960} in the
diffusive limit and numerically solve the Usadel equations \cite{Usadel1970}
for two and three dimensional weak link junctions.%
\begin{figure}
[t]
\begin{center}
\includegraphics[
height=1.228in,
width=3.3079in
]%
{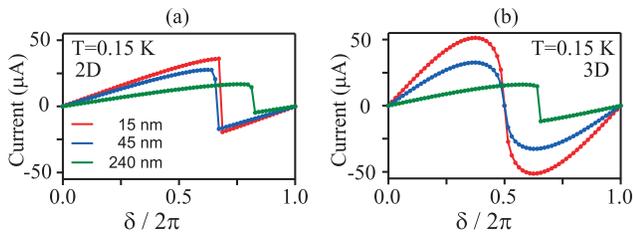}%
\caption{Computed current phase relations for an Al nanobridge with width
$W=45$ nm and three different lengths $L=15,45,$ and $240$ nm. Data are shown
for (a) 2D and (b) 3D banks at $T=0.15$ K. The 3D structures exhibit nonlinear
CPRs which progressively become linear as $L$ exceeds $\xi$. Corresponding
lengths for 2D structures exhibit a nearly linear, hysteretic CPR (the other
branch is suppressed for clarity). In both geometries, the maximum current
systematically decreases with increasing bridge length.}%
\label{fig_cpr}%
\end{center}
\end{figure}

We first compute the complex pairing potential $\Delta$ in the bridge and the
banks. The modulus and argument of $\Delta$ yield the superconducting gap and
phase, respectively. We use the $\Phi$ parametrization of the Usadel equations
\cite{Golubov2004} by making the substitutions $F=\frac{\Phi}{\sqrt{\omega
_{n}^{2}+|\Phi|^{2}}}$ and $G^{2}=1-|F|^{2}$ where $G$ and $\ F$ are the
normal and anomalous Green's functions. This leads to the equations
\begin{subequations}
\label{eqn_usadel}%
\begin{gather}
g(\omega_{n})(\Phi(\omega_{n})-\Delta)=\frac{D_{0}}{2}\mathbf{\nabla}\left[
g(\omega_{n})^{2}\mathbf{\nabla}\Phi(\omega)\right]  \label{eqn_usadel1}\\
g(\omega_{n})=\frac{1}{\sqrt{\omega_{n}^{2}+|\Phi(\omega,\mathbf{r})|^{2}}%
},\Delta=\frac{\sum_{n}g(\omega_{n})\Phi(\omega_{n})}{\sum_{n}\frac{1}%
{\sqrt{\omega_{n}^{2}+1}}}\label{eqn_usadel2}%
\end{gather}
where $\Delta$ and $D_{0}$ are the gap function and diffusion constant
respectively, rescaled by $\Delta_{0}$, the zero temperature value of the gap.
Finite temperature effects enter into the above equations through the spacing
$\Delta\omega=2\pi T/\Delta_{0}$ of the rescaled Matsubara frequencies
$\omega_{n}$. The Usadel equations are smooth in the frequency $\omega_{n}$
and hence the low temperature limit is expected to be reached when the
temperature is about 10 times smaller than the gap $\Delta_{0}=1.764k_{B}%
T_{C}\approx2.0$ K in Al. We find by comparing our numerical results in the
K-O limit that an upper cut-off frequency of $8.0$ K is sufficient for
convergence. The equations are numerically solved self-consistently for
$\Delta$ and $g$ on a 2D grid. For 3D structures, we use a cylindrically
symmetric geometry to reduce it to an effective 2D problem. The solutions on
the 2D grid are calculated for different values of superconducting phase
difference across the device and the current density is calculated using%
\end{subequations}
\begin{equation}
J(r)=\frac{\sigma}{e}\pi T\sum_{\omega}g(\omega)^{2}\text{Im}(\Phi^{\ast
}(\omega)\partial_{x}\Phi(\omega))\label{eqn_current}%
\end{equation}
where $\sigma$ is the conductivity of the metal film, $e$ is the electron
charge and the gradients are computed along the length of the bridge.%
\begin{figure}
[b]
\begin{center}
\includegraphics[
height=2.9827in,
width=3.2932in
]%
{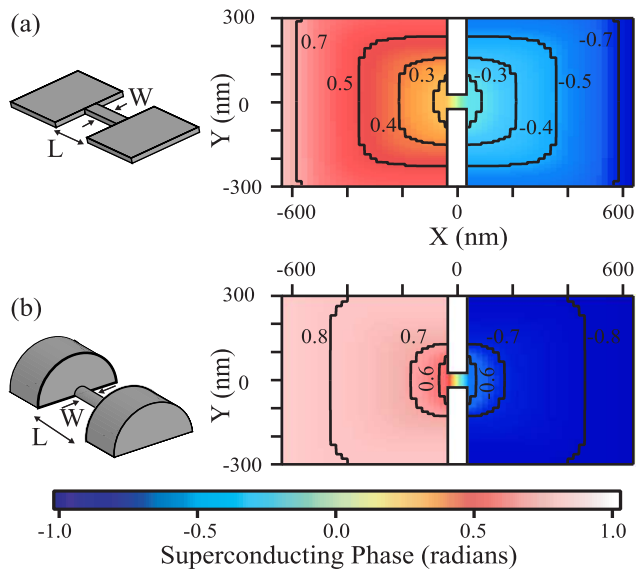}%
\caption{Superconducting phase (color) as a function of position for an Al
nanobridge with $L=75$ nm and $W=45$ nm. In (a) the banks are 2D with $600$ nm
lateral dimensions, and in (b) they are 3D as realized by a hemi-cylinder with
$600$ nm diameter. Contours of constant phase are also indicated. A total
phase difference of 2 radians is imposed symmetrically across the bridge. In
the 3D structure, most of the phase drop is across the constriction whereas in
the 2D planar geometry, there is significant phase variation in the banks.}%
\label{fig_2dvs3d}%
\end{center}
\end{figure}

The computed CPRs for an Al bridge with width $W=45$ nm and three different
lengths are shown in Fig. \ref{fig_cpr}. Panels (a) and (b) compare bridges
with 2D and 3D banks, respectively, at $T=0.15K.$ We take $T_{C}=1.2K$. The
banks for each bridge are $600$ nm in each lateral dimension. For the 3D case,
the banks also extend in the vertical direction as a hemi-cylinder of $600$ nm
diameter. For 3D structures, the CPR resembles a distorted sinusoid for the
shortest bridge length $L=15$ nm$\sim0.5\xi(0)$. As the bridge length exceeds
$\xi$, the current varies more linearly with the phase and the value of the
maximum current decreases. A similar trend is seen for 2D structures of the
same lateral dimension, but they exhibit a nearly linear CPR, even for
$L<\xi(0)$. To validate our numerical simulation, we compared our result for a
$15\times15$ nm bridge contacted to ideal phase reservoirs with the analytical
K-O result and found quantitative agreement. When computing the current for 2D
devices, the bridge cross-sectional area was chosen to match that of the
corresponding 3D devices by choosing the appropriate thickness for the 2D
device. Note that the CPRs in general should be an odd function of $\delta$
and also be $2\pi$ periodic which leads to the requirement that the current
should be zero at $\delta=\pm n\pi$ \cite{Golubov2004}. All the curves for the
2D device and that for $L=240$ nm for the 3D device have a non-zero value of
the current at $\delta=\pi$. However the full CPRs in these cases are
multivalued and the other branch (suppressed in Fig. \ref{fig_cpr} for
clarity) goes through zero at $\delta=\pi$.

In order to understand the different shape of the CPR in the 2D and 3D
geometries, we plot in Fig. \ref{fig_2dvs3d} the superconducting phase as a
function of position in a $75$ nm long, $45$ nm wide weak link
junction---dimensions readily achieved with e-beam lithography. For these
plots, a phase difference of 2 radians is imposed symmetrically at the ends of
the banks ($X=\pm637.5$ nm) and variation with position is calculated. The
phase is indicated by color and contour lines in Fig. \ref{fig_2dvs3d}. In the
3D case, the phase varies mostly in the bridge region and quickly heals to the
imposed value in the banks within a few $\xi(0)$. Thus, the banks act as good
phase reservoirs and the structure has a nonlinear CPR. In contrast, for the
2D case, the phase evolves both in the bridge and the banks, with a
logarithmic approach to the imposed value at the boundaries. Since the banks
fail to act as phase reservoirs, the entire structure resembles a
superconducting wire with a linear CPR rather than a Josephson junction. This
is particularly a problem when multiple 2D bridges are incorporated in a more
complicated circuit like a SQUID loop. There is significant overlap of the
weak link phases resulting in reduced modulation depth \cite{Hasselbach2002}.
Some of these limitations of 2D Dayem bridges have been discussed in
\cite{Likharev1979}. Variable thickness bridges are thus a more appropriate
replacement for tunnel junctions.%
\begin{figure}
[t]
\begin{center}
\includegraphics[
height=1.5065in,
width=3.3044in
]%
{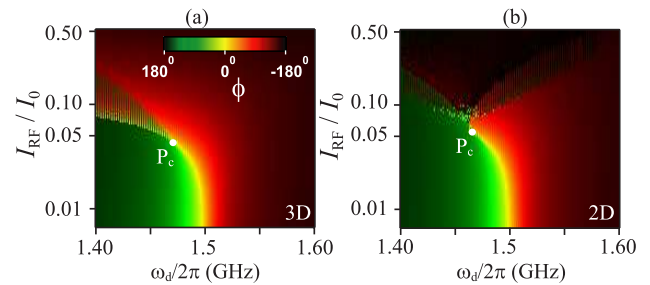}%
\caption{Microwave frequency response of a nonlinear oscillator constructed
using an Al nanobridge with $L=75$ nm, $W=45$ nm connected to (a) 3D and (b)
2D banks. The oscillator is driven with a single frequency $\omega_{d}$. The
phase $\phi$ of the steady state current oscillations, for a reflection setup,
is plotted as a function of $\omega_{d}$ and normalized drive current
$I_{RF}/I_{0}$. Increasing and decreasing current sweeps are interlaced to
highlight the hysteretic bistable region. Both devices exhibit pulling of the
resonant frequency (yellow) to lower values with increasing drive current,
typical of a driven nonlinear oscillator with a softening potential. However,
beyond the critical point $P_{c}$ (white dot), only the 3D device exhibits a
well defined bistable region (striped) where the oscillations are confined to
a single well of the periodic potential.}%
\label{fig_sims}%
\end{center}
\end{figure}

We now consider both classical and quantum anharmonic oscillators formed by
capacitively shunting a 3D weak link junction. We neglect the small intrinsic
capacitance of the nanobridges \cite{Likharev1979} and assume that the plasma
frequency is determined by the additional shunting capacitance only. For the
classical case, we compute the oscillating junction voltage under a microwave
drive. We consider a simple oscillator model where a weak link with
generalized CPR $I(\delta)=I_{0}f(\delta)$ with $\max\left\vert f(\delta
)\right\vert =1$ is shunted in parallel with a capacitance $C_{S}$ and a real
impedance $R_{S}$ which sets the quality factor of the oscillator. The
oscillator is driven with a time dependent current $I(t)=I_{RF}\cos\left(
\omega_{d}t\right)  $. The resulting equation of motion is
\begin{equation}
C_{S}\varphi_{0}\frac{d^{2}\delta(t)}{dt^{2}}+\frac{\varphi_{0}}{R_{S}}%
\frac{d\delta(t)}{dt}+I_{0}f(\delta(t))=I_{RF}\cos\left(  \omega_{d}t\right)
\label{eqn_drivenJJ}%
\end{equation}
The equation of motion was solved numerically using a fourth order Runge-Kutta
method for CPRs corresponding to nanobridges with $W=45$ nm, $L=75$ nm
contacted with 2D and 3D banks. The CPRs for these two devices differ in the
maximum current $I_{0}$ and the zero bias inductance $L_{J0}=\varphi
_{0}(\partial I(\delta)/\partial\delta)^{-1}|_{\delta=0}$. We chose values of
$C_{S}$ and $R_{S}$ to obtain a small oscillation resonant frequency of $1.5$
GHz and a quality factor $Q=50$ for both samples---typical parameters in
bifurcation amplifier circuits \cite{Siddiqi2004}. The results of the
simulation are shown in Fig. \ref{fig_sims}a,b. The steady state oscillation
phase, measured in reflection, is plotted in color as a function of drive
frequency $\omega_{d}$ and normalized drive current $I_{RF}/I_{0}$. Let us
consider the variable thickness bridge first. At low drive amplitude
$I_{RF}/I_{0}<0.01$, one recovers the familiar linear resonance behavior as
the phase evolves from $180$ to $-180$ degrees. The zero crossing of the phase
(yellow) corresponds to the resonant frequency. As one drives the oscillator
with larger drive current, the resonance frequency shifts to lower values as
is expected for a softening potential. For currents beyond a critical value
$P_{c}$, indicated as a white dot in the figure, the oscillator becomes
bistable and can coexist in two dynamical states (dashed region). This
behavior is typical of driven nonlinear oscillators \cite{LandauLifshitz1960}
and has been experimentally verified in Josephson tunnel junction based
oscillators \cite{Siddiqi2005}. This suggests that nanobridges with 3D banks
can be used for potential applications in the field of bifurcation and
parametric amplification \cite{Yurke1989}. On the other hand, we do not
observe a well defined bistable region in the microwave response of the Dayem
bridge structure. This is because accessing the nonlinear regime requires
larger phase excursions and the phase $\delta$ can hop between neighboring
wells of the periodic potential \cite{Manucharyan2007}. The critical point can
be pushed to lower drive currents but this requires significantly higher
values of Q.

Finally, we consider the possibility of constructing a quantum bit using weak
link junctions. Since the critical current of these junctions tends to be
large ($\sim20\mu A$), it is natural to consider a phase qubit geometry in
analogy with tunnel junction circuits with $\mu$A critical
currents\cite{Martinis2002}. We will only present results for a 3D\ weak link
junction and compare it with a Josephson tunnel junction since nanobridges
with 2D banks have multivalued CPRs which complicates calculations and the
potential operation of such a qubit. For the purpose of comparison, we also
numerically solve the Schr\"{o}dinger equation of capacitively shunted current
biased Josephson junctions and find the quantum bound state energy levels. We
choose a tunnel junction with a critical current equal to that of the weak
link junction. An appropriate shunting capacitance is used to fix the zero
bias plasma frequency of the tunnel junction device to $\omega_{P0}/2\pi=20$
GHz. A static current bias equal to $99\%$ of $I_{0}$ is used to tilt the
washboard potential to reduce the plasma frequency to about $7.5$ GHz, which
is typical for phase qubits \cite{Martinis2002}. The 3D bridge with $L=75$ nm
and $W=45$ nm is capacitively shunted to achieve the same reduced plasma
frequency under identical current bias.
\begin{figure}
[t]
\begin{center}
\includegraphics[
height=1.8568in,
width=3.3209in
]%
{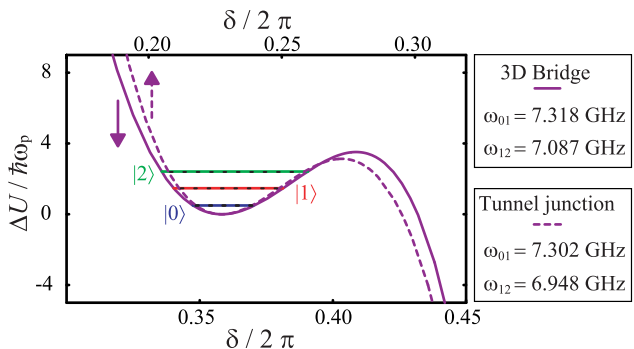}%
\caption{Quantum energy levels in a current biased, capacitively shunted
$L=75$ nm, $W=45$ nm Al nanobridge with 3D banks. For comparison, the
potential and energy levels of a tunnel junction with the same critical
current are shown as dashed lines. A current bias equal to $99\%$ of the
critical current is applied to obtain 3 quantum levels in the metastable
potential well. The shunting capacitances are chosen to yield identical plasma
frequencies. The level anharmonicity for the weak link junction ($\sim3\%$) is
only slightly smaller than that of the tunnel junction ($\sim5\%$).}%
\label{fig_quantumwells}%
\end{center}
\end{figure}

The potential wells and bound state energies are shown in Fig.
\ref{fig_quantumwells}. The tunnel junction data are shown as dashed lines.
The overall shape of the potential for the two cases is quite similar. The
positions of the minima occur at different values of $\delta$, as indicated by
two horizontal axes in the figure, since the maximum in the CPR of the tunnel
junction is at $\delta=\pi/2$ and at $\delta>\pi/2$ for the weak link
junction. The lowest bound states have nearly the same energy with the
subsequent higher levels differing by a few percent. For the tunnel junction,
$\omega_{01}=7.302$ GHz and $\omega_{12}=6.948$ GHz, corresponding to an
anharmonicity of $\sim5\%$. For the weak link junction, $\omega_{01}=7.318$
GHz and $\omega_{12}=7.087$ GHz, corresponding to an anharmonicity of
$\sim3\%$, which is only slightly smaller than that of the tunnel junction and
suggests the plausibility of weak link phase qubits. Moreover, it is possible
to increase the anharmonicity by using a slightly higher plasma frequency.

In conclusion, we have investigated the microwave transport properties and
quantum energy level structure of weak link Josephson junction oscillators,
focusing in particular on optimizing anharmonicity. We computed the CPR for 2D
Dayem and 3D variable thickness bridges by solving the Usadel equations for
different bridge dimensions. The 3D structures exhibit strong nonlinearity---a
consequence of the fact that the phase drop in these structures is confined
mainly to the weak link. Under microwave drive, these junctions should exhibit
bifurcation phenomenon for practical bridge dimensions and oscillator quality
factor. Additionally, under static current bias, the quantum energy levels
closely mimic ideal tunnel junction behavior, suggesting the possibility of
constructing phase qubits without any lossy oxide barriers, giving potentially
enhanced coherence. Nanobridge junctions proposed here can also potentially be
used to couple superconducting circuits to other solid state quantum systems
such as molecular magnets and quantum dots.

We would like to thank D.-H. Lee, B. Spivak, F. Wilhelm, R. Packard, M.
Hatridge and J. Clarke for useful discussions. R.V. and I. S. acknowledge
support from AFOSR Grant \#: FA9550-08-1-0104. J. D. S and M. L. C.
acknowledge support from NSF Grant No. DMR07-05941 and Director, Office of
Science, Office of Basic Energy Sciences, Materials Sciences and Engineering
Division, of the U.S. Department of Energy under contract No. DE-AC02-05CH11231.%

\end{document}